\documentclass[12pt]{article}   
\usepackage[pdftex]{graphicx}
\usepackage[pdftex]{color}
\usepackage{times}
\title{TRI-BN-15-15: Space Charge Limits in the Dae$\delta$alus DIC Compact Cyclotron}
\author{Rick Baartman, TRIUMF}
\date{December, 2015}
\begin{document}
\maketitle
\raggedright
\begin{abstract}
The Dae$\delta$alus DIC compact cyclotron is proposed to achieve an extracted current of $5$\,mA of singly-charged hydrogen molecules, even though unlike other compact cyclotrons, the extraction is not by stripping. The authors of the proposal consistently use perveance as a scaling argument that such high current is possible. The argument is shown to be incorrect and a realistic limit is calculated. This limit is likely about $0.2$\,mA and certainly no higher than $0.5$\,mA.\end{abstract}

\section{Introduction}
Compact H$^-$ cyclotrons have the nice feature that their phase acceptance is not limited by requiring all particles to have the same turn history. For example, the TR30 series of cyclotrons have phase acceptance of about $60^\circ$\cite{kuodevelopment}. Particles on the RF crest take $150$ turns, particles at phase acceptance extremes take $150/\cos 30^\circ=173$ turns. In principle, then, to reach $2$\,mA extracted from such a cyclotron would require as little as $12$\,mA peak current injected.

This is opposite to cyclotrons like the proposed DIC that require separated turns at extraction: these require all launched particles to stay together; all take the same number of turns, the same amount of acceleration time. Operation in this mode for TR30 would require to close the phase acceptance by a factor of $5$, reducing output by at least the same factor. In principle, flattening the waveform with additional harmonics can alleviate this, but it has only ever been achieved in separated sector (non-compact) cyclotrons.

The DIC is to deliver $5$\,mA of mass=$2$ (singly-charged molecular hydrogen) at 60 MeV/u \cite{winklehner2015rfq}. The phase acceptance is $5\%$ or $18^\circ$. This means the peak current is at least $100$\,mA. The proponents of the project make the claim (repeatedly, see \cite{calabretta2011preliminary,alonso2012high,barlow2013high,abs2012multimegawatt,axani2016high,campo2013high}) that the advantage of using higher mass particles is that the perveance is lower. This reasoning is incorrect in two crucial ways. The perveance they calculate is based on the average beam current of $5$\,mA. Space charge is a local effect, depending upon the peak current. The $100$\,mA peak current is an order of magnitude {\bf higher} than has ever been achieved in compact cyclotrons.

Secondly, the perveance of protons in compact cyclotrons is being compared with the perveance of H$_2^+$ in DIC  for the same $\beta$, not the same injection energy. Indeed then the perveance for equal peak currents is lower by a factor of 2 for H$_2^+$, but this ignores the fact that the transverse focusing that counteracts space charge is one quarter as large in DIC because the orbit radius is twice as large in the $1$\,Tesla central field. Instead of being twice as good, it is twice as bad for the same cyclotron field.

\section{Theory}
\subsection{2D Model}
The Kapchinsky-Vladimirsky equations can be written in terms of the perveance. For envelope radius $x,y$, tune $\nu$, orbit radius $R$,
\begin{equation}x''=-\frac{\nu_x^2}{R^2}x+\frac{\epsilon_x^2}{x^3}+\frac{2K}{x+y}\end{equation}
and similarly for $y$. The space charge limit (for zero emittance $\epsilon$; finite emittance is even worse) is thus given by the ratio of the perveance $K$ to the focusing force $(\nu/R)^2$. Whereas for fixed beam speed $\beta$, the perveance is indeed proportional to mass$^{-1}$, the proper scaling figure of merit is not $K$, but $R^2K$. Since for a given central field $B$, $R\propto$ momentum, and (again for fixed beam speed) this is proportional to mass.
 
Thus, the hydrogen molecules are a factor of 2 worse rather than a factor of 2 better than hydrogen ions. This knocks the expected DIC $5$\,mA down to $1.2$\,mA. Taking into account that the phase acceptance at $18^\circ$ is a factor $3$ smaller than achieved in H$^-$ compact cyclotrons, another factor of $3$ is lost. This predicts an upper limit of $0.4$\,mA for the DIC.

These arguments are very rough and moreover the use of continuous beam envelope equations breaks down when bunches are short. So let us turn to a 3D approach to estimate current limit.

\section{3D Model}
For the DIC, RF is $33$\,MHz and injection energy is $80$\,keV, therefore $\beta\lambda=80$\,mm. So this means the accepted bunch is $4$\,mm in length, making it shorter than the anticipated turn width. But at the space charge limit, the radial and longitudinal motions couple strongly eventually resulting in bunches that are precisely circular\cite{yang2010beam}. In other words, if the bunch is radially wider than it is long, it will in the space of a turn or so rotate to the point that it is longer than it is wide, too long compared with the phase acceptance needed for single turn extraction. Eventually after a few turns, because space charge is nonlinear, these propellering bunches will adopt and fill the circular shape. Thus to maintain the short bunches required for clean extraction, it is required that the radial width also be $4$\,mm.

The device used to bring the bunches onto the median plane of a compact cyclotron is an inflector. It strongly couples all three directions in phase space: there is no way to deposit a bunch that is much smaller radially than vertically. Therefore to deposit bunches that meet the phase acceptance criterion, their initial shape must be close to spherical.

Let us first examine the evolution of a spherical bunch in free space. 
Radius is $r$, charge is $Q$: the force equation for a boundary particle of charge $e$ is
\begin{equation}m\frac{d^2r}{dt^2}=\frac{1}{4\pi\epsilon_0}\frac{eQ}{r^2},\end{equation}\ \begin{equation}r''=\frac{1}{4\pi\epsilon_0mv^2}\frac{eQ}{r^2}\end{equation}
Initial radius $r_0$ will expand. What is the drift distance that causes it to double in size?
\begin{equation}z_{\rm double}=2.3\sqrt{r_0^3\frac{4\pi\epsilon_0}{Q}\frac{mv^2}{2e}}=2.3\sqrt{\frac{r_0^3}{c}\frac{V_0}{(30\Omega)Q}}\end{equation}
Insert DIC parameters: $r_0=2.1$\,mm, $V_0=80$\,kV, $Q=5{\rm mA}/33{\rm MHz}=150$pC.
\begin{equation}z_{\rm double}=54\,{\rm mm}\end{equation}

In order for space charge to be handled, this length ought to be large compared with the focal scaling lengths in the cyclotron. It's not.

The motion internal to the sphere is linear and yet the sphere size obeys a nonlinear evolution equation given above. This can be demonstrated for a uniformly filled sphere, simply using Gauss' law. Sacherer\cite{sacherer1970rms} demonstrated that the linear evolution is still correct for any realistic distribution of charge as long as the radius is replaced by twice the rms size.

As was done above for the 2D continuous beam model, We can find the space charge limit by comparing the space charge defocus force with the orbital focusing. Assume that the bunch is well matched to the smooth focusing that applies in the cyclotron at centre. Then the beam envelope size is constant. The space charge defocusing effect is 
\begin{equation}\label{e6}\left.r''\right|_{\rm space\ charge}=\frac{1}{4\pi\epsilon_0mv^2}\frac{eQ}{r_0^3}\ r\end{equation}
At the limit, this is balanced by the focusing force from the cyclotron field:
\begin{equation}\label{e7}\left.r''\right|_{\rm betatron}=-\frac{\nu^2}{R^2}\ r\end{equation} At injection, the tune $\nu=1$.
In fact, if eqns.\,\ref{e6},\ref{e7} sum to zero, there is no resultant force remaining to contain the beam. A careful self-consistent analysis \cite{baartman2013space} shows that the space charge force can be only as large as one half the focusing force. Thus
\begin{equation}\frac{1}{4\pi\epsilon_0mv^2}\frac{eQ_{\rm max}}{r_0^3}=\frac{1}{2R^2}\end{equation} Or, in simpler form,
\begin{equation}Q_{\rm max}=\frac{V_0}{30\Omega}\frac{r_0^3}{cR^2}\end{equation}

Let's insert numbers appropriate to the DIC. As stated, $r_0=2.1$\,mm; also $R=58$\,mm, $V_0=80$\,kV. We find $Q_{\rm max}=24$\,pC. At $33$\,MHz, this is $800\,\mu$A, a far cry from $5$\,mA, but even this is an overestimate. It requires that the bunches have zero emittance and are placed precisely on the first turn with exactly the correct correlation between energy and radial position and between radial divergence and longitudinal position to initiate the vortex motion. The inflector used in compact cyclotrons can do none of these things: far from it, the inflector has strong and unfortunate coupling between all 3 degrees of freedom. 

A hint can be gained from looking at the particular case of the TR30. This is known to run at $1$\,mA, usually runs at $0.75$\,mA \cite{cojocaruoperational}, but has shown output as high as $1.2$\,mA. $V_0=25$\,kV, $R=19$\,mm, $r_0\sim 1.6$\,mm, this gives 2.4\,mA for $73$\,MHz RF. This is about a factor 2 higher than has been achieved.

Similarly, we expect that the $800\,\mu$A DIC limit is also  a factor of at least 2 optimistic. 

\subsection{RFQ Injection?} 
\begin{figure}[p]\centering
\includegraphics[width=0.6\textwidth]{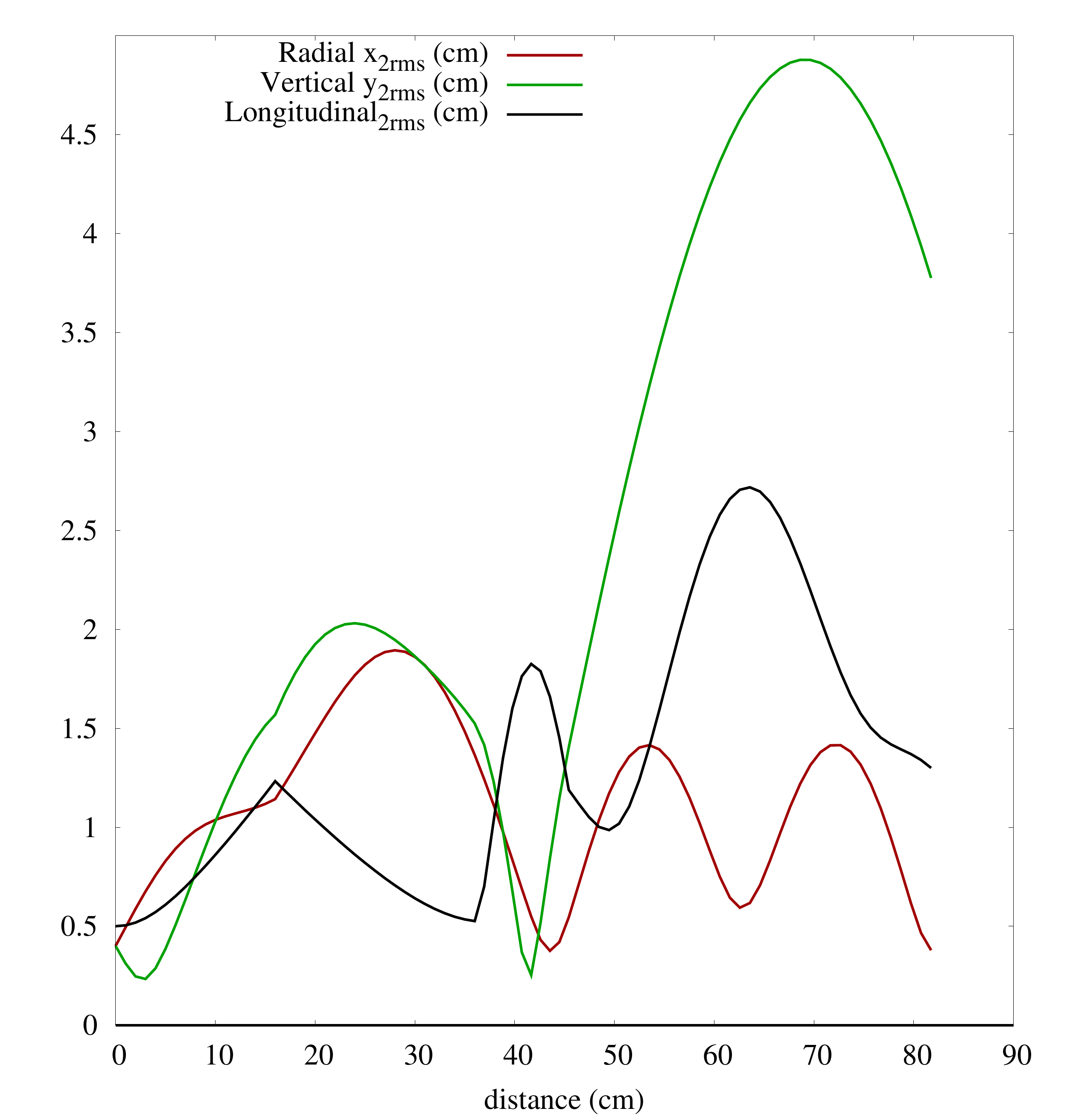}
\caption{Rebunching of RFQ beam. The buncher is an ideal thin lens at $16$\,cm, inflector entrance at $36$\,cm. The beam is transversely focused before the inflector by a solenoid that mimics the cyclotron's axial field. It has been adjusted to $0.9$\,Tesla to create best possible transverse focusing.}\label{f4}\end{figure}

As I show below (Fig.\,\ref{f4}), though the RFQ does a good job of bunching the beam, one needs an additional buncher because of the debunching between RFQ and inflector. 

Inflectors used in compact cyclotrons are highly dispersive. Roughly, the angular dispersion is $1$. This means that an energy spread of $dE/E=2dp/p=\pm 10\%$ \cite{winklehner2015rfq} generates vertical angles of $\pm 50$\,mrad. Optimistically, the vertical tune is $0.3$. In that case, beam vertical size becomes $\pm 0.05R/\nu\sim \pm R/6$ or $20$\,mm full size. This is in addition to size due to mismatch. It is the reason that existing compact cyclotrons use no or ``gentle'' bunching. No bunching means no energy spread; gentle bunching is a technique whereby the energy spread from the buncher becomes canceled by longitudinal space charge as the bunches drift to the inflector\cite{kuodevelopment}. The proposed RFQ solution for DIC injection\cite{winklehner2015rfq} creates a bunched beam. It is of no help; what is needed for the cyclotron is a gently bunching rather than already bunched beam.

So it is better off without an RFQ, since emittances will be smaller and energy spread ideally low. Another argument is that if the beam is unbunched, remaining DC, it can be space charge neutralized by collisions with the background gas generating electrons. In the case of H$^-$, this is highly beneficial.

\subsection{Matching}
The compact cyclotron has no way of matching the inflector output to the first turn optics. In fact, even aside from the unavoidable coupling of all three planes, there will be a strong vertical mismatch because the inflector has a much smaller vertical aperture than the matched beam size. At the very least, this loses another factor of 2. As well, the vertical focusing is much weaker than the radial, and moreover is phase-dependent since it arises from the RF: initial acceleration must occur on the falling side of the RF waveform because of the strong vertical defocusing on the rising side.

\subsection{Theory Conclusion}
\begin{figure}[p]\centering
\includegraphics[width=0.6\textwidth]{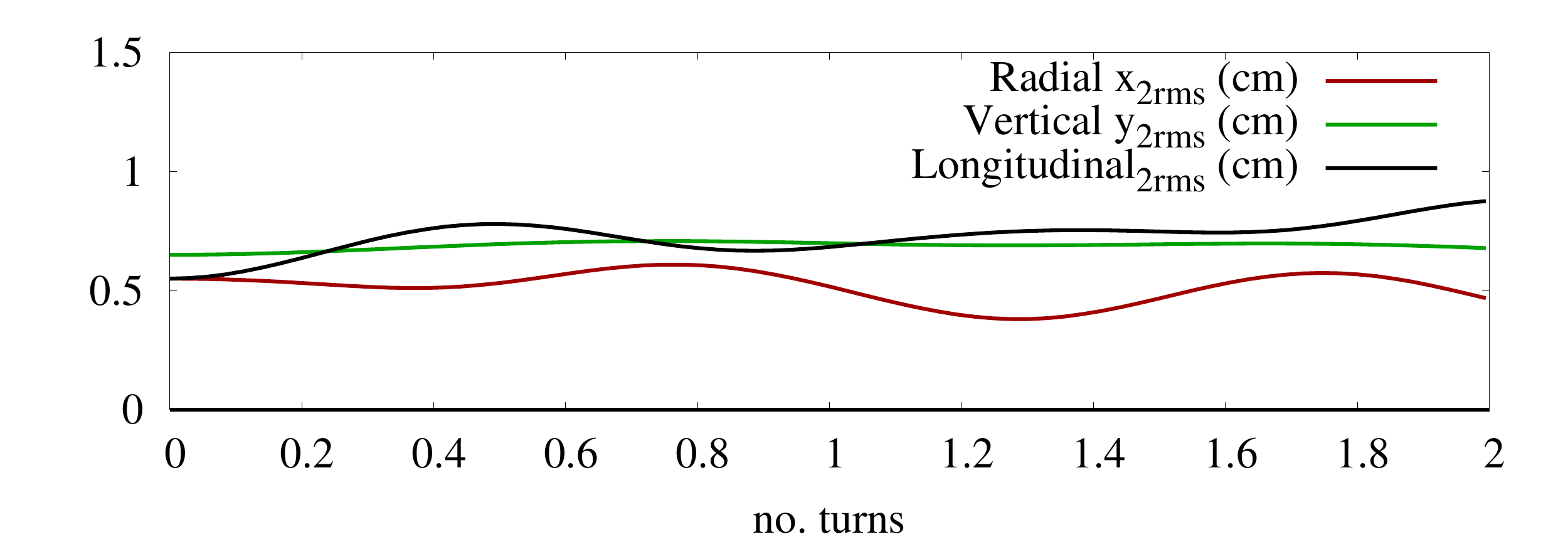}
\caption{Approximately-matched, zero emittance in all 3 planes.}\label{f7}
\end{figure}

I conclude that the DIC can accelerate no more than about $200\,\mu$A, a factor $25$ from the needed $5$\,mA. Increasing the bunch dimensions is also not an option. In order for the required charge $Q=150$\,pC to reach extraction energy, $r_0$ would need to be larger by a factor of 3, making the full $4\sigma$ size of the bunches $12$\,mm. See Fig.\,\ref{f7} for the envelope calculation. This is not possible for two reasons: (1) It's too large to fit through a reasonably-dimensioned inflector, and (2) the resulting bunch occupies $54^\circ$ RF. Because the vertical focusing comes from the RF,  particles would have energy gains ranging by a factor of $\cos(54^\circ)=0.6$. This is of no consequence when extraction is by stripping, but in the DIC machine, all particles must have same turn history.

\section{Envelope Evidence}
A good and simple way to check these results is to use an envelope code. {\tt TRANSOPTR} \cite{dejong1983first} can calculate the complete 3D bunch envelope evolution, with space charge; it tracks all 21 second moments. I have added routines to handle: the cyclotron's axial field, the inflector, and cyclotron dipole with energy gain. This was used to design the new vertical section of beamline for the TRIUMF cyclotron\cite{baartman2009cyclotronDN}. The new line was built, commissioned, and performs as predicted by the envelope calculations \cite{baartman2011IPAC2}.

\begin{figure}[p]\centering
\includegraphics[width=0.6\textwidth]{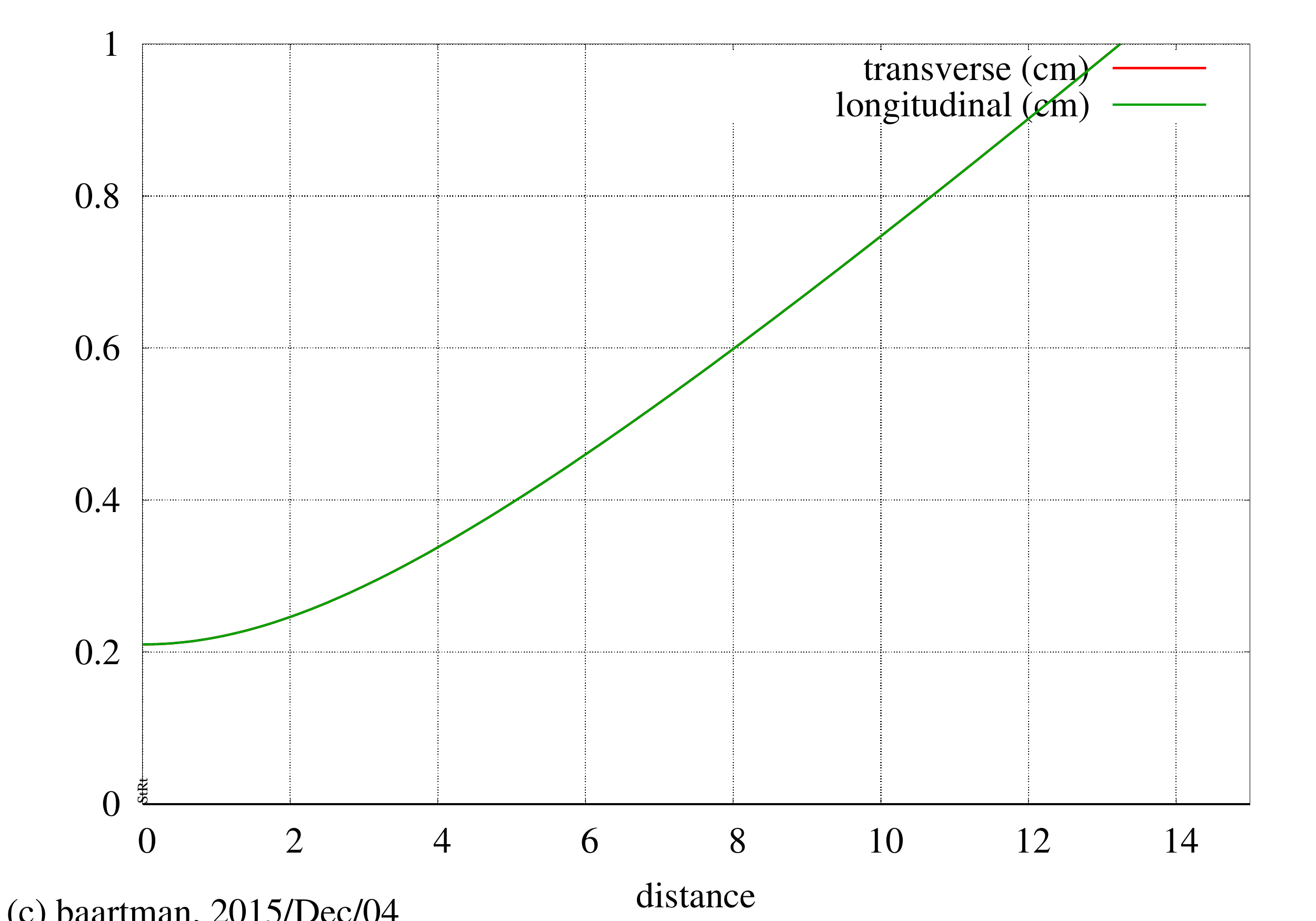}
\caption{Evolution of a spherical bunch of charge drifting in free space. $Q=150$\,pC, $80$\,keV hydrogen singly charged molecules.}\label{f1}\end{figure}

\begin{figure}[p]\centering
\includegraphics[width=0.6\textwidth]{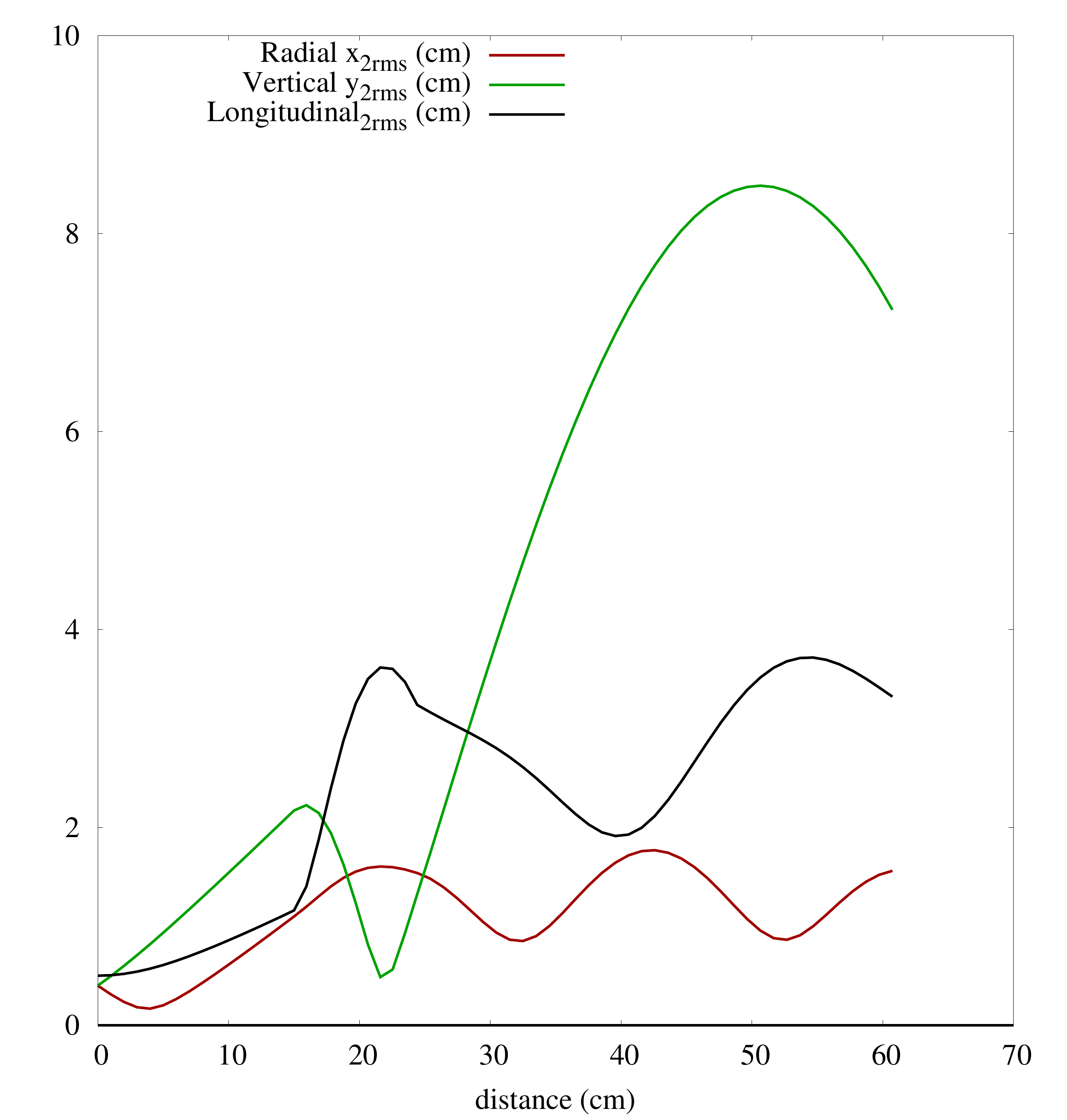}
\caption{Bunches from RFQ\cite{winklehner2015rfq} drifting $15$\,cm to and through inflector onto the cyclotron median plane. The inflector exit is at $25$\,cm, and bunch is tracked for a further $36$\,cm: 1 turn.}\label{f2}\end{figure}

Let us first as a check use {\tt TRANSOPTR} to calculate the drifting spherical bunch. Fig.\,\ref{f1} shows that indeed, the DIC bunch doubles in size after $54$\,mm drift.
Next, let us drift the beam from the RFQ\cite{winklehner2015rfq} into the inflector (Fig.\,\ref{f2}). The inflector was chosen to place the bunches near the first turn orbit: Height = $6$\,cm, magnetic radius $R=5.78$\,cm, tilt parameter = $-1$. The cyclotron field is given an index to mimic a vertical tune of $0.3$. In each of these plots, the bunch is tracked through a full first turn.

(Note that the evolution in the first $15$\,cm is substantially in agreement with Fig.\,5 of Winklehner et al.\cite{winklehner2015rfq}; their graph shows RMS sizes, and we show 2RMS sizes.) Clearly, this leads to absurdly large bunches. Keep in mind that in order for turn width to stay low enough for clean extraction, the black line should stay at or below $0.2$\,cm. This is violated by a factor of $20$.

\begin{figure}[p]\centering
\includegraphics[width=0.6\textwidth]{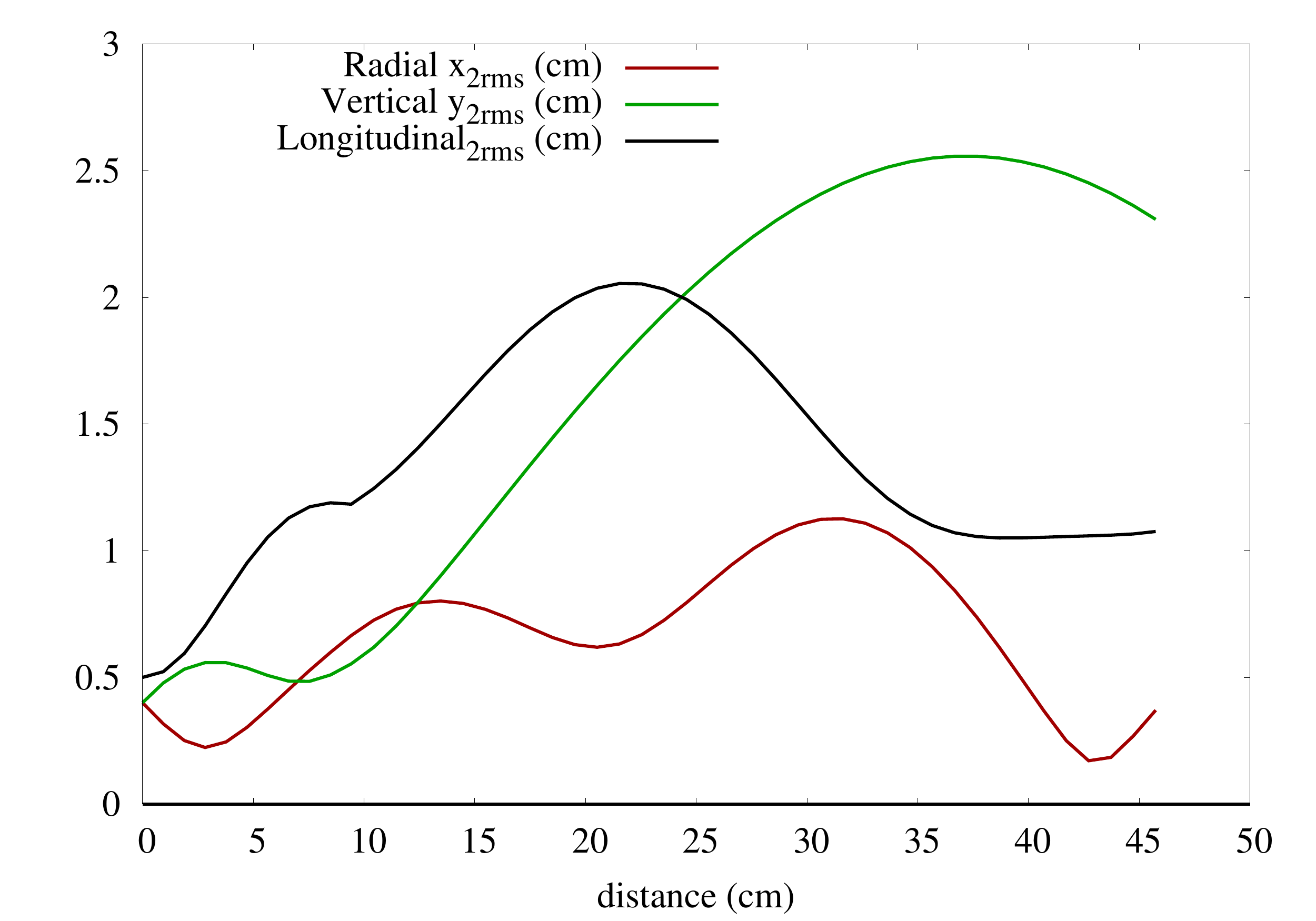}
\caption{As in previous figure, but the drift of $15$\,cm shortened to zero. The first $0$ to $10$\,cm is the inflector, then from $10$\,cm to $46$\,cm is one turn in the cyclotron.}\label{f3}\end{figure}

To reduce debunching, it would be better to reduce the distance to injection. If we (unrealistically) place the RFQ so that its exit coincides with the inflector entrance, we have the result in Fig.\,\ref{f3}.

Lastly, using the RFQ beam but re-bunching the beam to its minimum possible length (Fig.\,\ref{f4}).
In this case, the axial magnetic field is used to achieve a transverse focus in the inflector, so the bunches at least fit through it. We see that bunch radius $r$ is typically about $1$\,cm radially and $2$\,cm longitudinally, but up to $4.5$\,cm vertically. This huge size is partly due to mismatch and partly to dispersion. These sizes are incompatible with the size of the centre region magnet gap and RF dee electrodes.

\section{Ideally-Placed Bunches}
\begin{figure}[p]\centering
\includegraphics[width=0.6\textwidth]{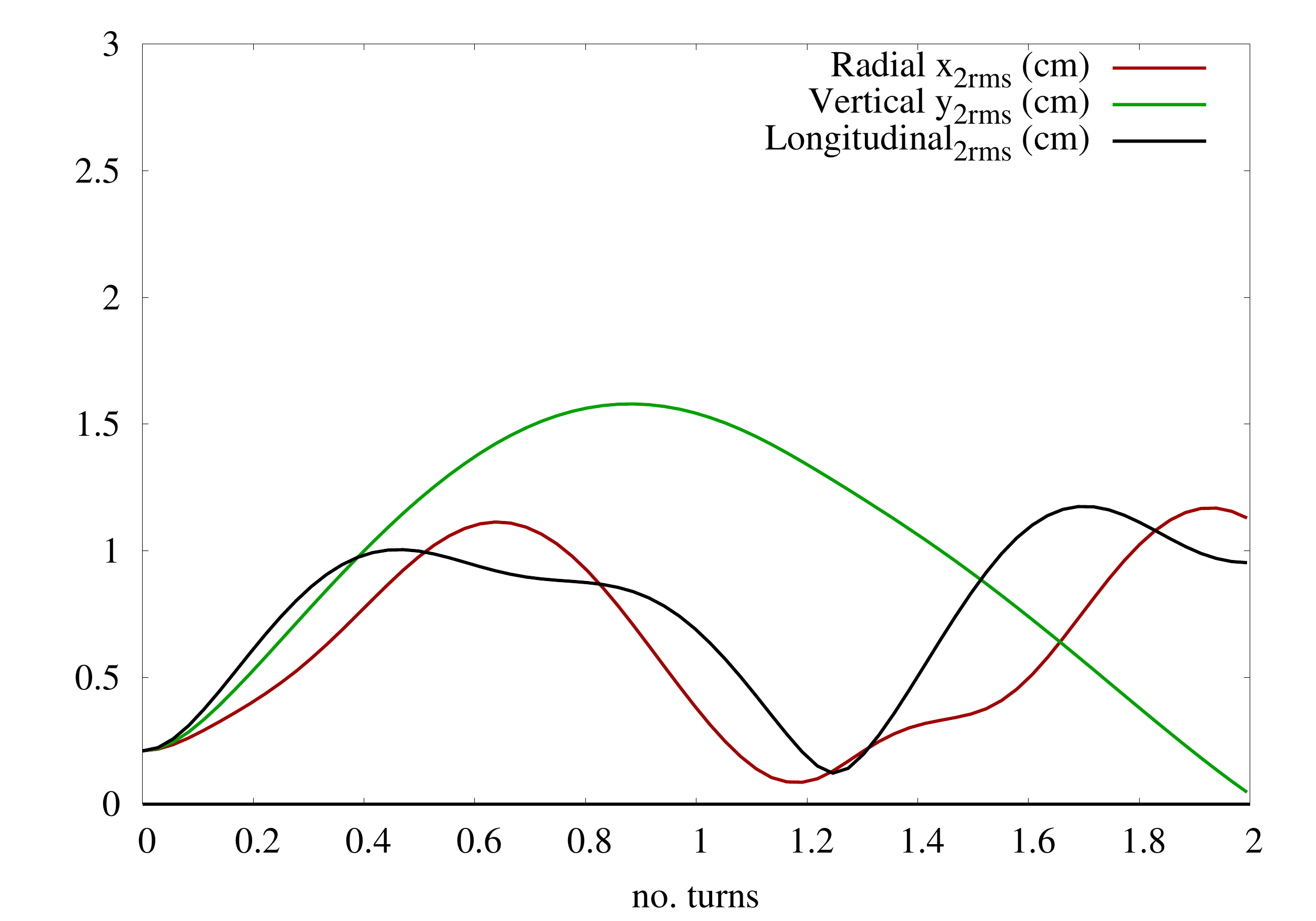}
\caption{Evolution of bunch of desired size; zero emittance in all 3 planes.}\label{f5}
\end{figure}

Ignoring the practicality of the inflector, if we place $r=2.1$\,mm zero emittance bunches on the first turn, the evolution is as Fig.\,\ref{f5}.

\begin{figure}[p]\centering
\includegraphics[width=0.6\textwidth]{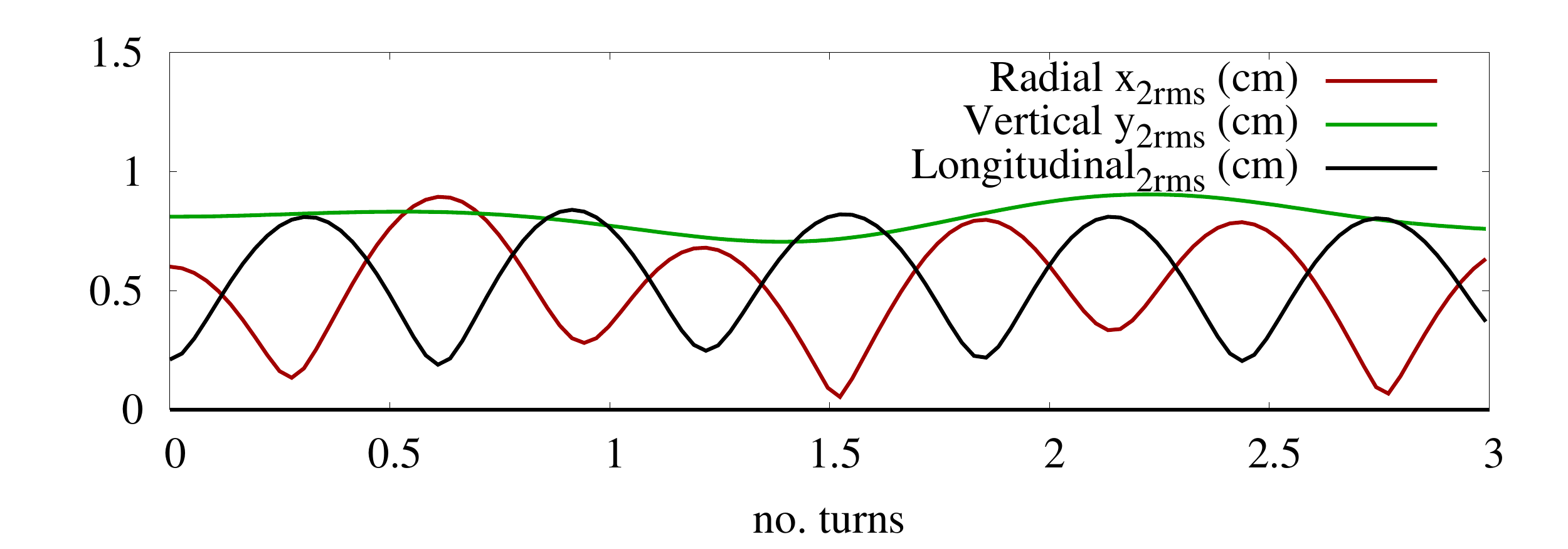}
\caption{Matched vertically but radial-longitudinal propellering; zero emittance in all 3 planes.}\label{f6}
\end{figure}

We can match vertically by increasing the size, but if we want to inject bunches as short as required while trying to match radially, the result is as Fig.\,\ref{f6}.
This is a propeller-like case, with radial and longitudinal swapping sizes. Realistically because of the nonlinearity of space charge, the bunches will smear to fill the swept area.

Lastly, we can somewhat match the bunch by making the dimensions sufficiently large (Fig.\,\ref{f7}). This is roughly $6$\,mm in radius as calculated in the Theory section; it is the smallest bunch that can contain the $150$\,pC of charge.

\section{Conclusions}
The proposal to extract $5$\,mA of H$_2^+$ without stripping from a compact cyclotron fails on two counts. (1) The requisite short bunch cannot be created by the injection system, and (2) even if it could, the focusing at the cyclotron centre is too weak to contain it.
\clearpage 
\bibliographystyle{plain} 
\bibliography{Baartman,BaartmanDN,Others}
\end{document}